# The anomalous pressure behavior of tangential modes in single-wall carbon nanotubes

Wei Yang[1]    Ru-Zhi Wang[1]*    Yu-Fang Wang[2]    Xue-Mei Song[1]    Bo Wang[1]    Hui Yan[1]*

[1]Laboratory of Thin Film Materials, Beijing University of Technology, Beijing 100022, China

[2]Department of Physics, Nankai University, Tianjin 300071, China

Using the molecular dynamics simulations and the force constant model we have studied the Raman-active tangential modes (TMs) of a (10, 0) single-wall carbon nanotube (SWNT) under hydrostatic pressure. With increasing pressure, the atomic motions in the three TMs present obvious diversities. The pressure derivative of $E_{1g}$, $A_{1g}$, and $E_{2g}$ mode frequency shows an increased value ($d\omega/dP > 0$), a constant value ($d\omega/dP \approx 0$), and a negative value ($d\omega/dP < 0$) above 5.3 GPa, respectively. The intrinsic characteristics of TMs consumedly help to understand the essence of the experimental T band of CNT. The anomalous pressure behavior of the TMs frequencies may be originated from the tube symmetry alteration from $D_{10h}$ to $D_{2h}$ then to $C_{2h}$.

To whom correspondence should be addressed: Tel: 86-01067392412. Fax: 86-01067392412
Email: wrz@bjut.edu.cn; hyan@bjut.edu.cn

In the case of carbon nanotubes (CNTs) at high pressure, their vibrational spectra probed by Raman spectroscopy have been found particularly fruitful both as a characterizational tool and a testing ground for the theoretical predictions.[1,2] In particular, Raman-active vibrational modes, such as radial breathing mode (RBM) and tangential modes (TMs), have attracted considerable attention,[2-15] because they are not only strongly resonance enhanced but also sensitive to the structural deformation of CNTs. The reduction in intensity and the broadening of R and T bands with increasing hydrostatic pressure have been observed in some investigations both for SWNT bundles[2-11] and individual SWNTs.[14] Several Raman studies reported a disappearance of the R band and a decreased pressure derivative (but still $d\omega/dP > 0$) of the T band near 2 GPa,[3-7] which was interpreted as the sign of a subtle structure transition. Moreover, Amer et al.[8] have observed a plateau ($d\omega/dP \approx 0$) of the T band under increasing pressure and have ruled out significant deformation of bundles below 10 GPa. Whereas, softening ($d\omega/dP < 0$) of certain TMs between 10~16 GPa has been observed by Teredesai et al.,[10,11] which may, although, be related to the solidification of the pressure-transmitting medium (PTM) at 10 GPa.[3,8] Clearly, a change in the pressure derivative ($d\omega_T/dP$) of the T band occurs under increasing pressure, but the exact nature of this change has proved controversial. Most authors seem to favor a change to the hexagonal,[4,5] oval[6] cross-section deformation or tube collapse,[14] while others propose an "adsorption"-like molecular ordering of the PTM around tubes.[8,9] However, in fact, differences in sample composition (e.g. diameters and chiralities), inevitably intertube interaction in the bundles and variances in experimental conditions, especially in terms of the PTM and the laser excitation power, may mask or smear the nature of the TMs change with pressure.[3] Therefore, the intrinsic characteristics of TMs for an isolated SWNT under hydrostatic pressure should be traced out, and it may consumedly help to understand the essence of TMs of CNT. With regard to TMs, corresponding to the characteristic A, $E_1$, and $E_2$ modes located around 1600cm$^{-1}$, they are experimentally difficult to distinguish from one another because of their similar frequencies.[16] Therefore, the experimentally observed T band is, in fact, composed of three Raman-acitve TMs, and the peak position is naturally used to refer to the T band.

Here, we will mainly focus on the effect of hydrostatic pressure on the TMs, e.g. $A_{1g}$, $E_{1g}$ and $E_{2g}$ modes, of a (10 ,0) SWNT. Our theoretical calculations were performed using the constant-pressure molecular dynamics (MD) simulation,[17] in which the interactions between carbon atoms are obtained by empirical Tersoff-Brenner potential.[18] The simulation time step is 1fs and the residual force per atom is 0.01 eV/Å in the structural optimization. The TMs including $A_{1g}$, $E_{1g}$ and $E_{2g}$ modes were calculated by combining above MD results with the force constant model.[15,19] The present scheme has been carefully checked and a detailed RBM study of isolated SWNTs under pressure has been published.[15] In this work, the anomalous pressure behavior of TMs are clearly observed that three TMs manifest distinct characteristics from one another above 5.3 GPa. For example, $E_{1g}$, $A_{1g}$, and $E_{2g}$ mode frequency shows an increased value ($d\omega/dP > 0$), a constant value ($d\omega/dP \approx 0$), and a negative value ($d\omega/dP < 0$) above 5.3 GPa, respectively. Moreover, a linear blueshift and a decreased pressure derivative under increasing pressure observed by S. Lebedkin *et al.*[14] may be rationalized in terms of Lorentzian fitting of a sum of three calculated TMs to the experimental T band.

The optimized structures of a (10, 0) SWNT under different pressures are shown in Fig. 1(a). Clearly, the pressure induces mechanically cross-section shape transition from a circle to a convex oval then to a non-convex oval shape. For a (10, 0) SWNT, TMs located around 1600cm$^{-1}$ correspond to the characteristic $A_{1g}$, $E_{1g}$ and $E_{2g}$ modes, which are all out-of-phase motions. From Fig. 1(b)-(d), apparently, in circumferential ($E_{2g}$, $A_{1g}$) and axial ($E_{1g}$) TMs, two of the three nearest neighbor atoms move in opposite directions perpendicular and along to the tube axis, respectively. Note that $E_{2g}$ and $A_{1g}$ modes have the same C-C bond stretching motions as well as C-C-C bonds bending motions, but differ in the relative phase of their C-atom displacements in the unit cell. Being tangential to the nanotube surface are particularly sensitive to the nanotube strain. Therefore, with the pressure elevated, the motions of three TMs present obvious diversities in terms of C-atom displacements and its amplitudes as shown in Fig. 1(b)-(d) and Fig. 2. Especially, there are distinct changes of the amplitudes at some special points corresponding to the long axis and the short axis of an oval shape in the deformed tube. Microscopic dependence of these motions on the structural deformation will be discussed in detail later.

Furthermore, $E_{2g}$, $E_{1g}$ and $A_{1g}$ modes frequencies of a (10, 0) SWNT under different hydrostatic pressures are calculated and plotted in Fig. 3(b)-(d), respectively. In order to illustrate a dependence of TMs frequency transition on the structural deformation clearly, the length of an oval shape long and short axes as well as calculated energy, as a function of applied pressure are also plotted in Fig. 3(a). The important result here is that at 5.5 GPa a pressure-induced structural transition occurs corresponding to circle-to-oval shape changes; nevertheless, at 5.3 GPa the long and short axes has been unequal to each other, which may indicate a start in structural transition in elevated pressure runs. Interestingly, such a subtle structural change at 5.3 GPa affect $E_{2g}$ mode greatly in contrast to $E_{1g}$ and $A_{1g}$ modes, and an obvious softening occurs in the range from 5.3 to 5.5 GPa as shown in Fig. 3(b). We think that the softening of $E_{2g}$ mode may originate from the symmetry alteration of (10, 0) SWNT from $D_{10h}$ to $D_{2h}$ point group as shown in Fig. 1(a).

Microscopically, the structural transition is driven by competition between compression and bending of a tube under pressure.[20] Below 5.3 GPa, a circular tube shrinks by reducing its radius, which mainly costs compressive strain energy. So the tube at 5.3 GPa still holds $D_{10h}$ symmetry like that of the original tube without any pressure. Above 5.3GPa, because it is easier to bend than to compress a tube, the tube begins to greatly cost bending strain energy to increase curvature. At a critical pressure (5.5 GPa), the tube transforms from an analogous circle to an anisotropic oval shape, and then the symmetry lowers, with only 3 twofold rotational axes, to $D_{2h}$ point group according to carefully analysis of the new structural data. As the tube continues to shrink, it no longer compresses (maintaining its perimeter) but only bends (reducing its overall curvature), and then must adopt a shape to minimize bending energy. This eventually leads to another shape transition from a convex oval to a non-convex oval shape at 6.1 GPa. Meanwhile, the tube symmetry continues to lower, absent inversion centers, down to $C_{2h}$ group as shown in Fig. 1 (a). It should be noted that as the applied pressure elevates up to 7.0 GPa, the length of the short axis approaches nearly 3.35Å, which means that an additional van der Waals (vdW) interaction may lead to the collapse of the tube,[21] and then the TMs of the SWNTs may not be exhibited. Here, therefore, we only focus on the TMs characteristics of the (10, 0) SWNT subject to compression up to 6.8 GPa without regarding the vdW interaction.

Therefore, due to the pressure-induced symmetry alteration, not only the C-atom displacements (Fig. 1(b)-(d)) and its amplitudes (Fig. 2) but the frequencies (Fig. 3(b)-(d)) of the three TMs present obvious diversities. Especially, interestingly enough, the frequencies exhibit anomalous pressure behavior at certain critical pressure 5.3, 5.5, and 6.1 GPa. For example, $E_{2g}$ mode shows obvious softening between 5.3-5.5 GPa as shown in Fig. 3(b). Beyond 5.5 GPa, $E_{2g}$ mode frequency sharply shifts to higher frequencies, and then softens again slightly at 6.1 GPa; in terms of $E_{1g}$ mode (Fig. 3(c)), there is an increased pressure derivative between 5.5-6.1 GPa and then reaches a plateau; whereas $A_{1g}$ mode frequency (from Fig. 3(d)) deviates a constant dependence on the applied pressure between 5.5-6.1 GPa, and then softens considerably. For unambiguous evidence that the anomaly is an intrinsic property of CNTs,[3] the anomalous pressure behaviors of TMs, e.g. the softening or a plateau, would most likely arise from the change of phonon deformation potentials, attributed to structural deterioration in the radial direction, softening of the C-C intratubular bonds. It is important to note that these anomalous pressure behaviors of the three TMs are hardly observed experimentally because these TMs are difficult to distinguish from one another due to their similar frequencies. We hope that our results will inspire to carry out subtle experimental studies to identify the different TMs, once purified aligned samples become available.

Altogether, at low pressure, the three TMs all fit to linear equations and the fitted values of pressure derivatives, e.g. $d\omega_{E2g}/dP$ =5.8 cm$^{-1}$/GPa, $d\omega_{E1g}/dP$ =5.0 cm$^{-1}$/GPa, and $d\omega_{A1g}/dP$ =6.3 cm$^{-1}$/GPa, are given in the Fig. 3(b)-(d), respectively. These theoretical results are approximately agreed with the experimental value ($d\omega_T/dP$ =6.4 cm$^{-1}$/GPa) of individual disperse SWNTs.[14] Furthermore, at higher pressure, we predict that if the three TMs are Lorentzian fitted to the experimental T band which is identified with the peak position of a triplet of TMs ($A_{1g}$, $E_{1g}$, $E_{2g}$),[22] the change in $d\omega_T/dP$ of the fitted T band may be in accordance with experimental results.[14] Because the intensities of $A_{1g}$, $E_{1g}$ and $E_{2g}$ modes differently vary with increasing applied pressure[13] even with changes of experimental conditions, e.g. the PTM and the laser excitation power. And the peak position of the fitted T band is mainly based on the more intense mode. Thereafter, if the intensity of the softening mode (e.g. $E_{2g}$ mode in Fig. 3(b)) is weak, the fitted T band will exhibit no softening but a positive value of $d\omega_T/dP$ or a constant like the majority of experimental results.[3-9,14] Consequently, it becomes rational that the values of $d\omega_T/dP$ are different in diverse experiments under higher pressure. In addition, from Fig. 3(b)-(d), it is obvious that at distinct pressure (e.g. 5.3, 5.5 or 6.1GPa) the anomalous behaviors of three

TMs are presented, which would most likely make the transition of the fitted T band uncertain. Therefore, we agree with some authors[8,14] that the plateau or a change in the pressure derivative of the T band cannot reliably be interpreted as a sign of the structural transition, whether our prediction is true or not. A theoretical Lorentzian fitting in order to conceivably confirm the experimental result,[14] combined the acquired frequencies with coming calculated resonant Raman intensity under a series hydrostatic pressures, is planned for the future.

In summary, we investigate the intrinsic characteristic features in three TMs of an isolated (10, 0) SWNT under hydrostatic pressure. The results show that the atomic motions in the three TMs present obvious diversities with increasing pressure. The most interesting finding is the anomalous pressure behavior of different TMs frequencies, for example, $E_{1g}$, $A_{1g}$, and $E_{2g}$ mode frequency shows an increased value ($d\omega/dP > 0$), a constant value ($d\omega/dP \approx 0$), and a negative value ($d\omega/dP < 0$) above 5.3 GPa, respectively. The intrinsic characteristics of TMs for an isolated SWNT under hydrostatic pressure consumedly help to understand the essence of the experimental T band of CNT. These anomalous behaviors may be originated from the tube symmetry alteration from $D_{10h}$ to $D_{2h}$ then to $C_{2h}$ with increasing pressure.

**Acknowledgement**

This work is supported by PHR (IHLB), the National Natural Science Foundation of China (No. 10604001, No.60576012), and the Natural Science Foundation of Beijing (No. 4073029). We would like to thank Professor G. X. Cheng for useful discussions also.

**Figures captions：**

**FIG.1** (a) The cross-section shape elevation of (10, 0) SWNT at some selected pressures 5.3, 5.5, 5.8, 6.1 and 6.8 GPa, respectively, and corresponding symmetry group is attached. (b)-(d) The atomic motions of $E_{1g}$, $A_{1g}$ and $E_{2g}$ modes at symmetry transition pressures 5.3, 5.5 and 6.1 GPa, respectively.

**FIG.2.** The calculated amplitudes of half a unit cell atoms arranged in a line for (10, 0) SWNT at some selected pressure 5.3, 5.5, 5.8, 6.1 and 6.8 GPa, respectively.

**FIG.3.** (a) The energy as well as the length of the long and short axes, as a function of pressure for (10, 0) SWNT. (b)-(d) the calculated frequencies vs. pressure of $E_{2g}$, $E_{1g}$ and $A_{1g}$ mode, respectively, and the pressure derivatives mentioned alongside.

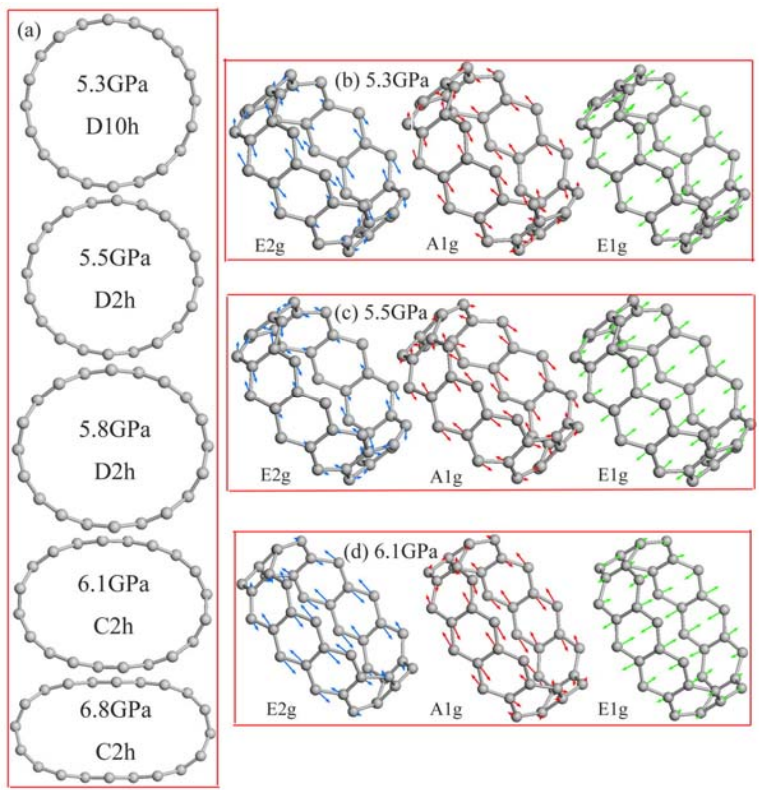

**FIG. 1        Yang et al.**

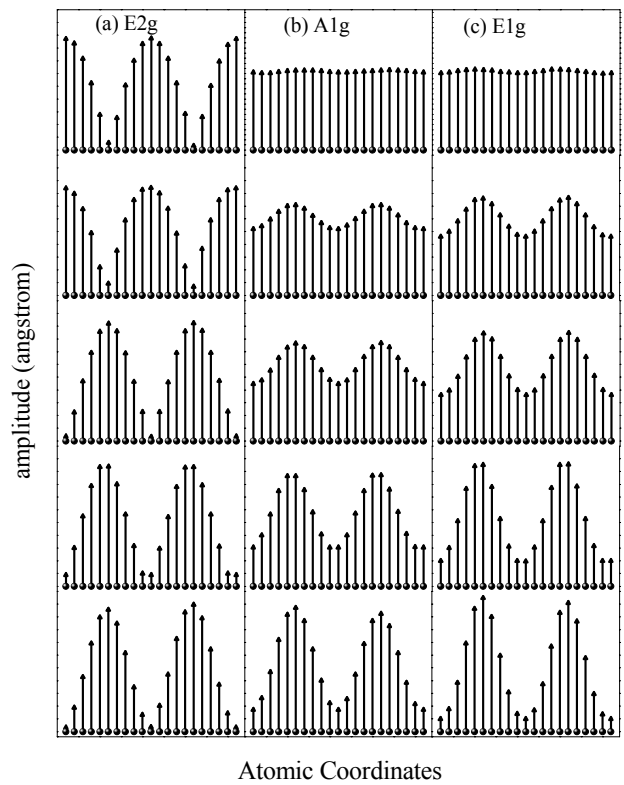

**FIG. 2**     **Yang et al.**

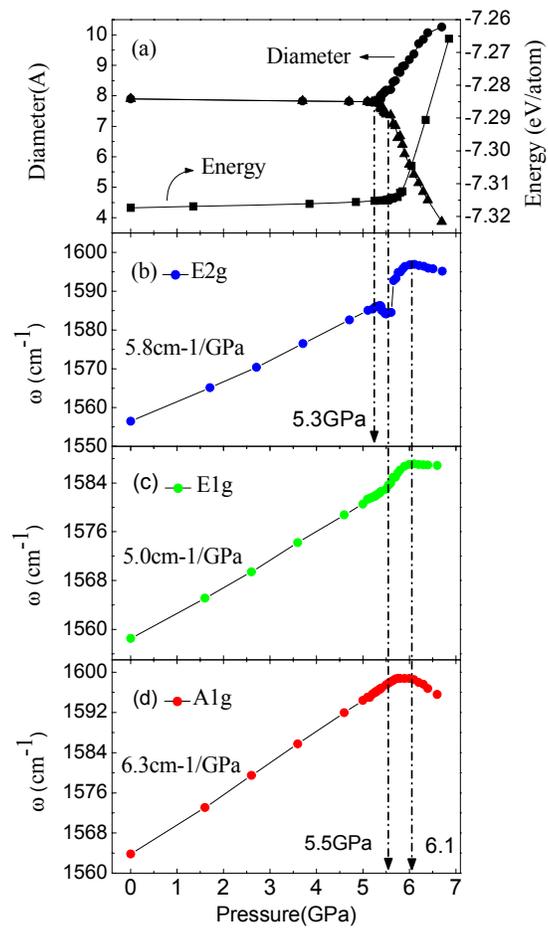

**FIG. 3      Yang et al.**